# Direct observation of chiral edge current at zero magnetic field in odd-layer MnBi$_2$Te$_4$

Jinjiang Zhu[1†], Yang Feng[2†], Xiaodong Zhou[3,4,5†], Yongchao Wang[6†], Zichen Lian[7], Weiyan Lin[1,3], Qiushi He[1], Yishi Lin[1], Youfang Wang[1], Hongxu Yao[1], Hao Li[8,9], Yang Wu[9,10], Jing Wang[1], Jian Shen[1,3,4,5,11,12], Jinsong Zhang[7,13,14]*, Yayu Wang[7,13,14], Yihua Wang[1,12]*

[1]*State Key Laboratory of Surface Physics and Department of Physics, Fudan University, Shanghai 200433, China*
[2]*Beijing Academy of Quantum Information Sciences, Beijing 100193, P. R. China*
[3]*Institute for Nanoelectronic Devices and Quantum Computing, Fudan University, Shanghai 200433, China.*
[4]*Zhangjiang Fudan International Innovation Center, Fudan University, Shanghai 200433, China.*
[5]*Shanghai Qi Zhi Institute, Shanghai, China*
[6]*Beijing Innovation Center for Future Chips, Tsinghua University, Beijing 100084, P. R. China.*
[7]*State Key Laboratory of Low Dimensional Quantum Physics, Department of Physics, Tsinghua University, Beijing 100084, P. R. China*
[8]*School of Materials Science and Engineering, Tsinghua University, Beijing, 100084, P. R. China.*
[9]*Tsinghua-Foxconn Nanotechnology Research Center, Department of Physics, Tsinghua University, Beijing 100084, P. R. China.*
[10]*Department of Mechanical Engineering, Tsinghua University, Beijing 100084, P. R. China.*
[11]*Collaborative Innovation Center of Advanced Microstructures, Nanjing, China.*
[12]*Shanghai Research Center for Quantum Sciences, Shanghai 201315, China.*
[13]*Frontier Science Center for Quantum Information, Beijing 100084, P. R. China*
[14]*Hefei National Laboratory, Hefei 230088, P. R. China*

† These authors contributed equally to this work.
* To whom correspondence and requests for materials should be addressed. Emails: jinsongzhang@tsinghua.edu.cn, wangyhv@fudan.edu.cn

**Abstract**

**The chiral edge current is the boundary manifestation of the Chern number of a quantum anomalous Hall (QAH) insulator. Its direct observation is assumed to require well-quantized Hall conductance, and is so far lacking. The recently discovered van der Waals antiferromagnet MnBi$_2$Te$_4$ is theorized as a QAH in odd-layers but has shown Hall resistivity below the quantization value at zero magnetic field. Here, we perform scanning superconducting quantum interference device (sSQUID) microscopy on these seemingly failed QAH insulators to image their current distribution. When gated to the charge neutral point, our device exhibits**

**edge current, which flows unidirectionally on the odd-layer boundary both with vacuum and with the even-layer. The chirality of such edge current reverses with the magnetization of the bulk. Surprisingly, we find the edge channels coexist with finite bulk conduction even though the bulk chemical potential is in the band gap, suggesting their robustness under significant edge-bulk scattering. Our result establishes the existence of chiral edge currents in a topological antiferromagnet and offers an alternative for identifying QAH states.**

**Introduction**

The bulk-boundary correspondence of topological phases dictates protected metallic states at the boundary of a bulk topological order [1–4]. For example, the surface of a three-dimensional topological insulator hosts surface Dirac cone states [5]. The existence of such surface Dirac bands serves as direct evidence for the bulk topology [6,7] even if the chemical potential does not lie in the bulk band gap, so that the bulk carriers play a part in the charge transport and relaxation of surface carriers [8]. For a 2-dimensional (2D) insulator, the nature of the edge state is a direct manifestation of the topological invariant of the bulk. A quantum spin Hall insulator [9–12] conserves time-reversal symmetry (TRS) and hosts helical edge states [13–15]. When TRS is broken in the bulk and proximitized with an *s*-wave superconductor, the edge is predicted to host exotic excitations useful for topological quantum computation [16–18].

A quantum anomalous Hall (QAH) insulator is an interesting topological phase that breaks TRS [19,20]. Its edge hosts the chiral edge state (CES), which represents the Chern number characterizing the TRS-broken topological invariant. In principle, along the edge of a macroscopic sample, such CES must propagate either clockwise (Fig. 1a) or counterclockwise (Fig. 1b), depending on the direction of magnetization. Due to the chirality as a consequence of broken TRS, CES cannot be localized by weak disorder [21], thus suppressing backscattering [22]. Such a one-dimensional fermionic state carries a current along the perimeter of the sample. In spite of the crucial role of CES in not only QAH but also the closely related quantum Hall effect, unlike its TRS counterpart, it has not been directly observed so far. Magnetic imaging experiments to detect chiral edge current have revealed complicated physics in the quantum Hall [23] and even a conflicting picture for QAH insulators [24].

A key consequence of CES in the QAH order is the quantized Hall conductance at zero field [25–27]. For a small enough bias on the leads, CES carries a chiral edge current, which ideally leads to quantized Hall conductance equal to $e^2/h$, where $e$ is the electron charge and $h$ is the Planck constant, even at zero magnetic field. Quantized anomalous Hall conductance was first observed in a magnetically doped topological insulator [25]. As this material required a delicate balance between a large magnetization and a low initial carrier doping, the quantization temperature was limited by magnetic disordering due to the randomly distributed magnetic dopants [28–30]. The edge current was absent, and instead current transport was bulk-dominant within the QAH regime of the doped system [24]. Furthermore, recent work nullified the earlier observation of edge Majorana fermions when proximitized by a superconductor [31]. Magnetic disorder is considered detrimental to the robustness of QAH and, therefore, possibly CES.

The recent discovery of a van der Waals antiferromagnet, MnBi$_2$Te$_4$, with a topological band structure brings hope of a clean material host for quantized Hall conductance and CES [32,33]. MnBi$_2$Te$_4$ consists of stacked Te-Bi-Te-Mn-Te-Bi-Te septuple layers (SL) and displays an A-type antiferromagnetic (AFM) order in which the magnetic moments of Mn order ferromagnetically within each SL and AFM between neighboring SL. Under a large magnetic field, odd-SL MnBi$_2$Te$_4$ thin flakes showed quantization temperatures up to a few tens of kelvin [34–36]. However, at zero magnetic field, the reported quantization was either not exact [37] or, in most cases, far from it [34–36,38–40]. Compared to the exfoliated MnBi$_2$Te$_4$ thin flakes, it is even harder for molecular beam epitaxy-grown MnBi$_2$Te$_4$ thin films to get close to the quantized regime, not to mention QAH at zero field [41–44]. Serious concerns were raised about the nature of the electronic state and magnetic order of MnBi$_2$Te$_4$ thin flakes at zero field, especially the fate of QAH. It is thus essential to elucidate the relationship between the current flow, CES, and the quantum transport in an intrinsic QAH.

**Detecting chiral edge current under an alternate-current bias**

In this work, we study odd-SL MnBi$_2$Te$_4$ thin flakes with a finite net magnetization at zero field (Fig. 1c). The 7-SL sample is gate-tuned to its charge-neutral regime by applying a back-gate voltage of $V_\mathrm{g} = 50$ V. Consistent with many previous reports [34–36,38–40], our transport measurement of the sample at the base temperature of 1.7 K yields a small anomalous Hall resistivity $\rho_{yx} \sim 820\ \Omega$ after magnetizing the sample under a 9 T field, seemingly suggesting a failed QAH (Fig. S8). In order to image the current distribution under weak perturbation, we employ the high flux sensitivity of scanning superconducting quantum interference device (sSQUID) microscopy. As a local magnetometer, sSQUID images flux through its pickup loop of 2 μm in diameter in our experiment [45]. Such magnetic flux has two sources: one from the nonzero net magnetization of 7-SL MnBi$_2$Te$_4$, and the other from the applied bias current through the device. We apply an alternate current (AC) with a small amplitude for lock-in detection of current flux to distinguish the magnetic field generated by the applied current from that of the magnetization.

The lock-in detection of chiral edge current under an AC bias is quite different from that of non-chiral current. First, suppose we apply a pure AC bias current without any direct-current (DC) offset and bias the device from the right side into a QAH with clockwise CES (Fig. 1d). The current flows along the bottom edge from right to left in the first half-cycle (Fig. 1e). During the second half, the chemical potential on the right electrode becomes lower than the one on the left (Fig. 1f), and therefore the current flows on the top edge from left to right according to the Landauer-Büttiker picture [46]. The lock-in amplifier adds a $\pi$ phase shift to the flux signal in the second half-cycle, reversing the sign of the current flux on the top edge. Consequently, the edge currents will appear to be along the same direction (Fig. 1g), and the overall current flow is consistent with the direction of the applied current. By adding a DC offset to the AC bias, we overcome the limitation of the lock-in detection of CES. For a non-chiral current transport, a DC offset has no effect on the current signal demodulated by the lock-in amplifier. The situation is quite different for CES. When the DC offset is larger than the AC amplitude (Figs. 1h-k), the current flows on the bottom edge for the whole period since the chemical potential on the right electrode is constantly higher than on the left. The $\pi$ phase shift to the flux signal subtracts the contribution from the second half-cycle (Fig. 1h, green shaded area), and we obtain current flux on the bottom edge proportional to $I_{AC}$ (Fig. 1k). (Using a negative DC offset leads to a current on the top edge, but the chirality will be reversed from its real one. Fig. S13)

We perform sSQUID microscopy of the chiral edge current of the MnBi$_2$Te$_4$ sample following the detection scheme we have discussed (Figs. 1i-l). We magnetize the sample under 9 T or -9 T out-of-plane magnetic fields at the base temperature and then remove the field for the measurements. The magnetizations are opposite using opposite fields on the 7-SL part but show little contrast on the 10-SL flake because of the AFM interlayer coupling (Figs. 2a and b). At the charge neutral point (CNP) of $V_g = 50$ V, we apply an $I_{DC} = 1$ μA offset to $I_{AC} = 500$ nA current bias on the right or left electrodes for both magnetizations to obtain four different current flux images (Figs. 2c-f). According to

Biot-Savart's law, the out-of-plane magnetic field generated by a current stripe has opposite signs on the two sides of the stripe, and the field is zero at the center. The 10-SL is conductive at $V_g = 50$ V and the current flows through it evenly, with its center part appearing white. The current flux does not change there as magnetization is reversed (Figs. 2c and d, or Figs. 2e and f). The sign of the flux exactly reverses when the direction of the current bias is reversed (Figs. 2c and e, or Figs. 2d and f). The DC offset in the bias also does not change the current flux of 10-SL (Fig. S14), which is consistent with non-chiral transport.

The current flow in the 7-SL flake is distinctively different from that in the 10-SL flake. For positive magnetization and biasing on the right electrode (Fig. 2c), the current flows predominantly along the lower edge and the boundary between the 7-SL and 10-SL. Meanwhile, there is little current flux contrast on the upper edge, even though it has a shorter path than the one the current takes. On the other hand, biasing from the left electrode with the same magnetization leads to current flowing on the upper edge clockwise (Fig. 2e). This is an indication of edge current breaking TRS in an odd-SL-layer sample at charge neutral. This chirality of the edge current switches upon changing the magnetization. Biasing from the right (left) electrode with negative magnetization, the current flows from the upper (lower) edge anti-clockwise (Figs. 2d and f). Overall, the chirality of the edge current is determined by the magnetization, whereas the edge that the current passes through depends on both the chirality and the source of the current bias. The current flux patterns under the four configurations are in complete agreement with our model for lock-in detection of chiral edge current (Fig. 1). These pieces of evidence unequivocally show the existence of CES at zero field despite the poor Hall quantization.

**Edge-bulk scattering of chiral edge current**

In the following, we investigate the influence of the bulk on the CES by varying the gate voltage. We fix the magnetization of 7-SL to be negative (Fig. 2b). At $V_g = 0$ V, the Fermi level is within the bulk conduction band, and the current is predominantly carried by the bulk (Fig. 3a). Reversing the bias direction reverses the current flux signal (Fig. 3b), which is consistent with the bulk current flow being non-chiral. Taking a line-cut across the 7-SL sample (Figs. 3a and b, dashed lines), we map out the continuous evolution of the current distribution as a function of $V_g$ (Figs. 3c and d). For the right-electrode bias, the current distribution is mostly bulk-centered below $V_g = 30$ V, at which point the center of the current starts to shift towards the top edge (Fig. 3c). The center reaches the top edge at $V_g = 50$ V and then turns back towards the middle upon further increasing $V_g$. The overall trend is similar for biasing from the left electrode (Fig. 3d). The center of the current shifts towards the bottom edge instead. The $V_g$ dependence of the current distribution rules out the bulk anomalous Hall effect as a cause for the chiral current flow at CNP. Instead, it shows that the observed chiral edge current originates from the in-gap CES in 7-SL MnBi$_2$Te$_4$, while outside the bandgap the current is non-chiral and carried by the bulk states.

Aware of any possible non-chiral current in the CNP regime, we quantitatively analyze the current distribution between the edge and the bulk in order to compare it with the resistance measurements. We assume the current in the four configurations (Figs. 2c-f) as the sum of two components $I_{u(l)}^{+(-)} + I_n$, where the first term is the chiral edge current (chirality +/-) either on the upper edge ($I_u$) or the lower edge ($I_l$) and the second term is the non-chiral bulk current. We obtain the edge-only chiral current flux (Fig. 3e) by taking the difference of the current flux images with opposite chirality but the same bias direction (Figs. 2c and d). This difference image cancels the $I_n$ component and leaves the flux contribution only from a clockwise chiral edge current: $I_l^+ - I_u^- = I_l^+ + I_u^+ = I^+$. By implementing the current reconstruction (SOI) from the difference image (Fig. 3e), we obtain the current density distribution, which circles around the boundary of the 7-SL flake (Fig. 3f). Alternatively, we can take the difference between current flux images from the left bias (Figs. 2e and f), and the result is similar. The sum (Fig. 3g) of the current flux images with the same bias but opposite chirality (Figs. 2c and d) has three components: $I_l^+ + I_u^- + 2I_n$. This sum image is very similar to the one obtained under zero DC offset (Fig. S14a), which is expected to be $(I_l^+ + I_u^-)/2 + I_n$ under the same $I_{AC}$ according to our earlier analysis (Figs. 1d-g). The reconstructed current density from the sum image exhibits all three components (Fig. 3h). All these observations support the coexistence of chiral edge and non-chiral bulk currents at CNP.

In order to separate the non-chiral current from the chiral current at CNP, we take line profiles from the difference and sum images. The line profile from the difference flux image is unipolar (Fig. 3i, orange), which could only be a result of circulating current. And the line profile from the difference current density image shows peaks corresponding to the upper and lower edges (Fig. 3j, orange). Meanwhile, the sum current density exhibits a broad spread (Fig. 3j, green) in agreement with its flux pattern (Fig. 3i, green) and the magnitude is larger than that of the chiral edge current density (Fig. 3j, orange) everywhere on the sample (Fig. 3j, gray region). However, there are still two

(though less distinctive) peaks in the sum current density (Fig. 3j, green), which distinguishes it from a uniform current ribbon. After subtracting the difference current density line profile from the sum and then dividing by two, we obtain a pure contribution from the non-chiral current $I_\text{n}$ (Fig. 3j, black). The line shape of $I_\text{n}$ is heavily centered in the middle, with little distribution on either edge. Furthermore, it has a different shape from the current density from the bulk carriers (Fig. 3j, purple), which is reconstructed from the current flux image at 0 V (Fig. 3a). This suggests that the non-chiral current at CNP has a different origin from the conventional bulk current outside the gap. After integrating the respective current density over the width of the line-profile, we find that about 70% of current flows along the upper and lower edges as chiral current, while the remaining 30% is non-chiral and non-edge.

The coherence length of ballistic chiral edge transport is much smaller than the size of a realistic open conductor. Historically, a confining potential for the edge current was proposed, which limits both elastic and inelastic scattering of edge carriers to be within the edge channel so that the quantization of Hall conductance could still be possible [22]. Nevertheless, longitudinal scattering within the edge channel is not enough to explain significant bulk conduction in the bulk exchange gap, and edge-bulk scattering is necessary. On the other hand, recent experiments suggest $MnBi_2Te_4$ suffers from strong chemical and magnetic disorders due to its metastable phase character [42,44,47–49]. One consequence of such disordering is the spatial fluctuation of the density of state (DOS) at the Fermi level, as reported in several scanning tunneling microscopy studies. Because the DOS at the Fermi level are mainly from the Bi-Te $p$ orbitals, such fluctuation implies local variations of the Bi and Te orbitals, which would affect the magnetic exchange interaction, leading to spatial variation of the magnetic exchange gap $\Delta(r)$. Such inhomogeneity assists the electron hopping from the edge to the bulk, especially in the regions where $\Delta(r)$ diminishes. Our bulk resistance measurement by shunting all the edge current to the ground [15] supports this scenario (Fig. 4a). When $V_\text{g}$ is tuned to CNP (Fig. 4b), we find that the voltage drop over the 7-SL peaks at around 4 mV while the bulk current reaches a minimum of 2 nA. The finite conductance at CNP suggests that when the bulk carriers are depleted (see SOI for more quantitative analysis), the current still has limited flow through the bulk, which is consistent with the bulk-edge scattering (Fig. 4c). As the external magnetic field increases, $\Delta(r)$ increases with it which effectively suppresses the edge-bulk scattering. Considering this scenario, we develop a realistic model to compare with the transport experiment.

We construct a Landauer-Büttiker (LB) multi-terminal model [46] to quantitatively understand the chiral transport under finite edge-bulk scattering (see supplementary S3 for details of the simulation). In the LB approach, one uses the transmission matrix elements $T_{j,i}$ to quantify the transmission probability between two electrodes *i* and *j* and, therefore, $I_i = \frac{e^2}{h} \sum_j (T_{j,i} V_i - T_{i,j} V_j)$ is the total current passing through the electrode *i*. To account for the chirality of a ballistic edge channel, one usually assigns 1 and 0 to $T_{j,i}$ and $T_{i,j}$, respectively, where electrodes *i* and *j* are the two nearest neighbors, and sets others to 0. Here we modify this method to also include the edge transport dissipation due to finite edge-bulk scattering. Specifically, we allow non-zero transmission matrix elements between any two terminals of the device because electrons can hop through the bulk between them. Moreover, instead of either 0 or 1, the value of $T_{j,i}$ can vary between 0 and 1, which are determined in a self-consistent way to reflect both the dissipation due to the edge-bulk scatterings as well as the chiral character of the edge conduction. In our multi-terminal model, except for the existing real voltage probes of the device, we also manually add *N* virtual floating probes between the neighboring real ones (Fig. 4d). Therefore, the dimension of the conductance matrix increases with *N*. Adding such virtual probes effectively increases the probability of edge-bulk scatterings because electrons will experience multiple scatterings among these virtual probes as they transmit from one real electrode to the other. Each scattering event has contributions from the bulk, i.e., electrons are scattered from the edge to the bulk and then back to the edge.

When the scattering probability is small (*N* = 2), the current is mostly carried by the edge states within the bulk exchange gap. But as expected for adding bulk scattering to ballistic chiral edge transport, the bulk current is small but non-zero. The effect of having a small edge-bulk scattering on the corresponding longitudinal resistivity is significant, which deviates from zero at CNP (Fig. 4e, blue circles). The simulated gate-dependent longitudinal resistance $\rho_{xx}(V_g)$ for different *N* from 2 to 12 (Fig. 4e) shows that in the CNP regime the system undergoes a transition from dissipationless to dissipative transport as *N* increases. At *N* = 12, the simulated $\rho_{xx}(V_g)$ (Fig. 4e, orange circles) matches the experimental data exactly (Fig. 4e, orange line), showing a peak at CNP and small resistance outside the gap. For $\rho_{yx}(V_g)$, the simulation also quantitatively agrees well with the experimental data in CNP and $V_g = 0$ (Fig. 4f). Some discrepancies appearing right outside the Zeeman exchange gap are possibly due to the global coexistence of QAH and the metallic domains when Fermi level $E_F$ is in the intermediate mixed region between gap and valence/conduction band with spatial fluctuation, which is likely illustrated by a percolation picture rather than our simple model of only considering edge-bulk scattering at the edges. Our model simulation demonstrates that significant edge-bulk scatterings play a major role in the quantization breakdown of the odd-SL sample. A reduced edge-bulk scattering, achievable through a more homogenous ferromagnetic exchange gap, may lead to better $\rho_{yx}$ quantization and suppress the dissipation.

**Conclusion**

We have observed a chiral edge current in odd-layer MnBi$_2$Te$_4$ at zero field, confirming the topological nature of its bulk electronic state, i.e., such CES comes from the topological nontrivial band structure. However, the finite bulk conduction and edge-bulk scattering jeopardize the transport quantization of such a QAH state. Our work establishes the chiral edge current as a more robust feature than the quantized transport to identify a QAH phase.

**Acknowledgement**

YHW would like to acknowledge support by the National Key R&D Program of China 2021YFA1400100, NSFC Grant No. 11827805 and 12150003, and Shanghai Municipal Science and Technology Major Project Grant No. 2019SHZDZX01. Y.F. acknowledges support by the NSFC Grant No. 11904053, the National Postdoctoral Program for Innovative Talents (Grant No. BX20180079), and the China Postdoctoral Science Foundation (Grant No. 2018M641904). X.Z. acknowledges support by the NSFC Grant No. 12074080 and 12274088. J.S. acknowledges support from National Key R&D Program of China (Grant No. 2022YFA1403300). The authors acknowledge Qian Niu for very helpful discussions on the anomalous Hall effect.

Figures

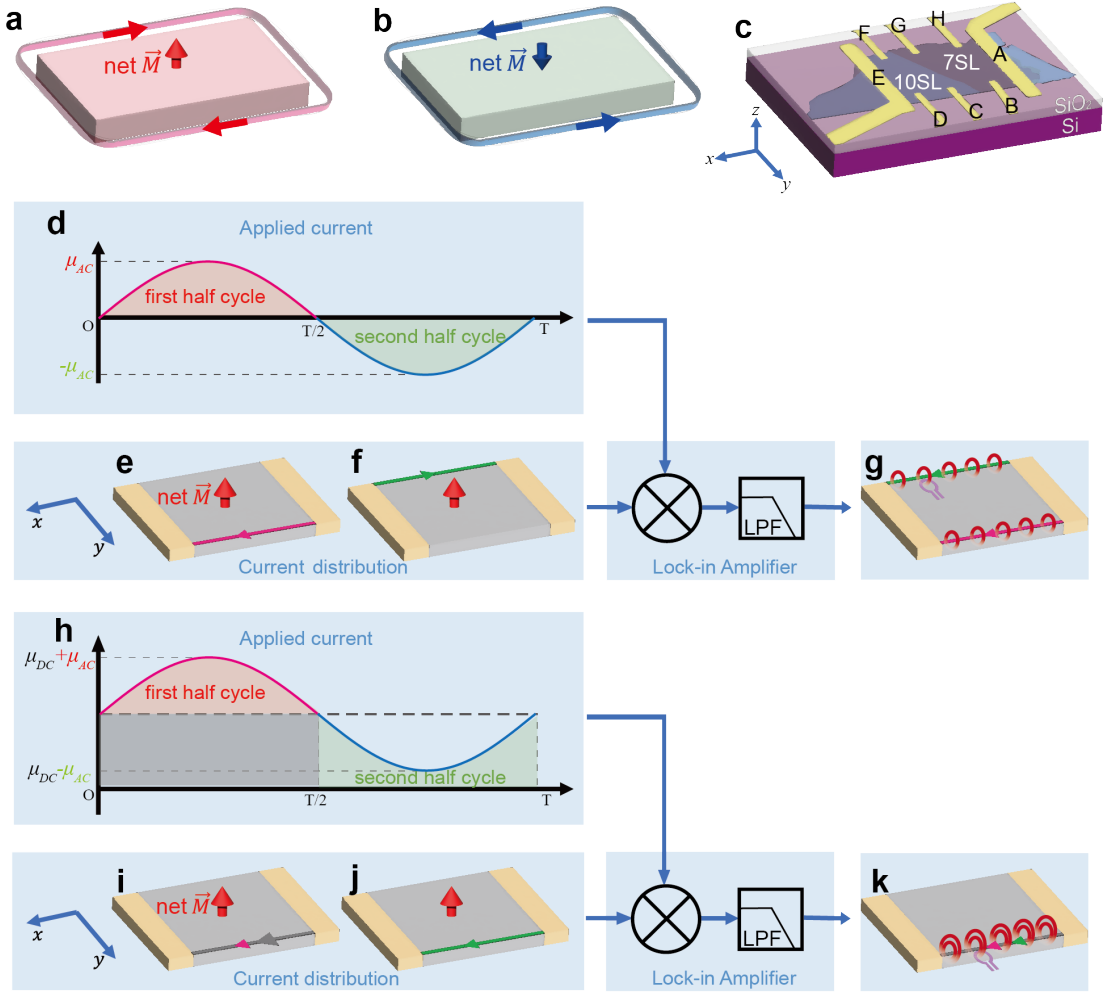

**FIG. 1 Chiral edge current detection scheme under an alternate-current bias**. **a** and **b**, Illustration of the direction of chiral edge current in a quantum anomalous Hall (QAH) insulator. The chirality of the edge current is clockwise and counter-clockwise under opposite out-of-plane magnetization *M*, respectively. **c**, Optical image of a MnBi$_2$Te$_4$ device consisting of a 7-septuple layer (SL) which potentially hosts QAH. There is also a 10-SL flake that has no net *M* and zero Chern number. **d**, Waveform of a pure alternate-current (AC) bias applied to QAH with period *T*. It has positive (negative) chemical potential $\mu_{AC}$ relative to ground during the first (second) half cycle. **e** and **f**, Current flow in the first and second half of the cycle, respectively, under the AC current bias in (**d**). **g**, A lock-in amplifier adds a $\pi$ phase shift to the current flux signal from the second half-cycle, which makes the currents from the top and bottom edges appear to be propagating in the same direction. **h-k**, Illustration of the situation when a direct-current (DC) offset is added to the AC bias. If the DC offset is larger than the AC amplitude, the chemical potential is positive during the entire period, and the current flows unidirectionally on the bottom edge. The strength of the demodulated flux signal is proportional to the amplitude of the AC amplitude and independent of the DC offset (twice the area of the pink region).

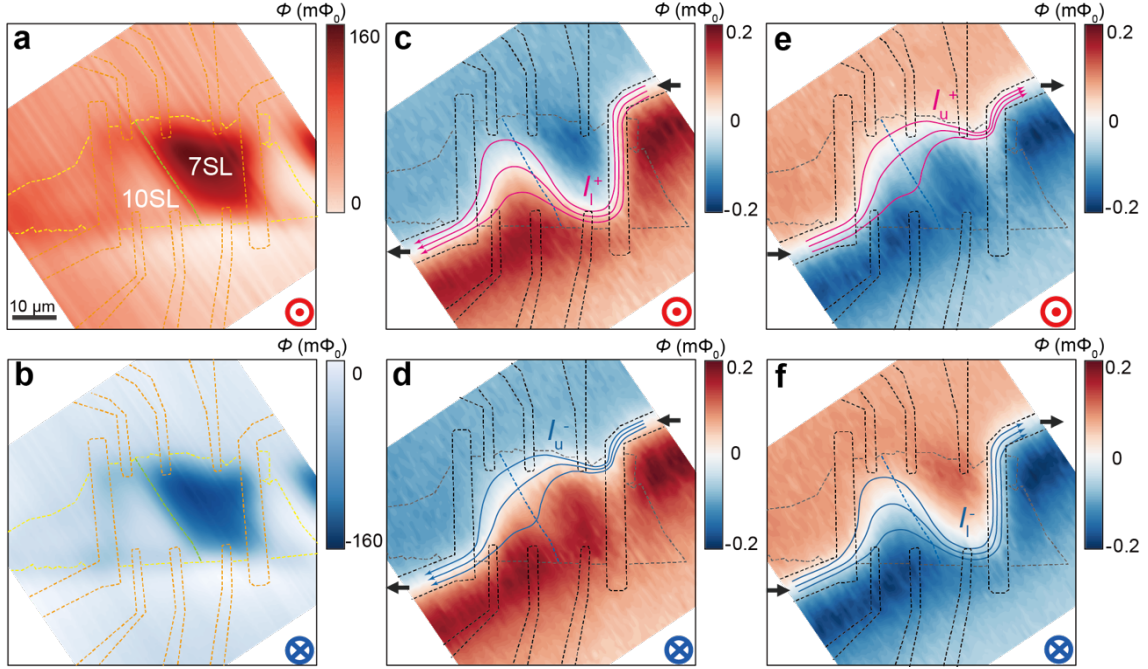

**FIG. 2 Chirality of the 7-SL edge states under opposite magnetization. a** and **b,** Static magnetic flux images after the sample is magnetized under 9 T (**a**) and -9 T (**b**), respectively. The 7-SL region shows a net *M*. **c** and **d,** Current flux images corresponding to the magnetization in **a** and **b**, respectively. A 500 nA AC current plus a 1 μA DC offset is applied to the right electrode. **e** and **f,** Current flux images with the same current bias as in **c** and **d** but applied to the left electrode. All the current flux images are taken at $V_g$ = 50 V. The streamlines are current flows around the 7-SL edge reconstructed from the current flux images. They show the edge current ($I_l$ for the lower edge and $I_u$ for the upper edge) and its chirality (magenta/+ for clockwise and blue/- for counter-clockwise), which predominantly carry the current under each configuration.

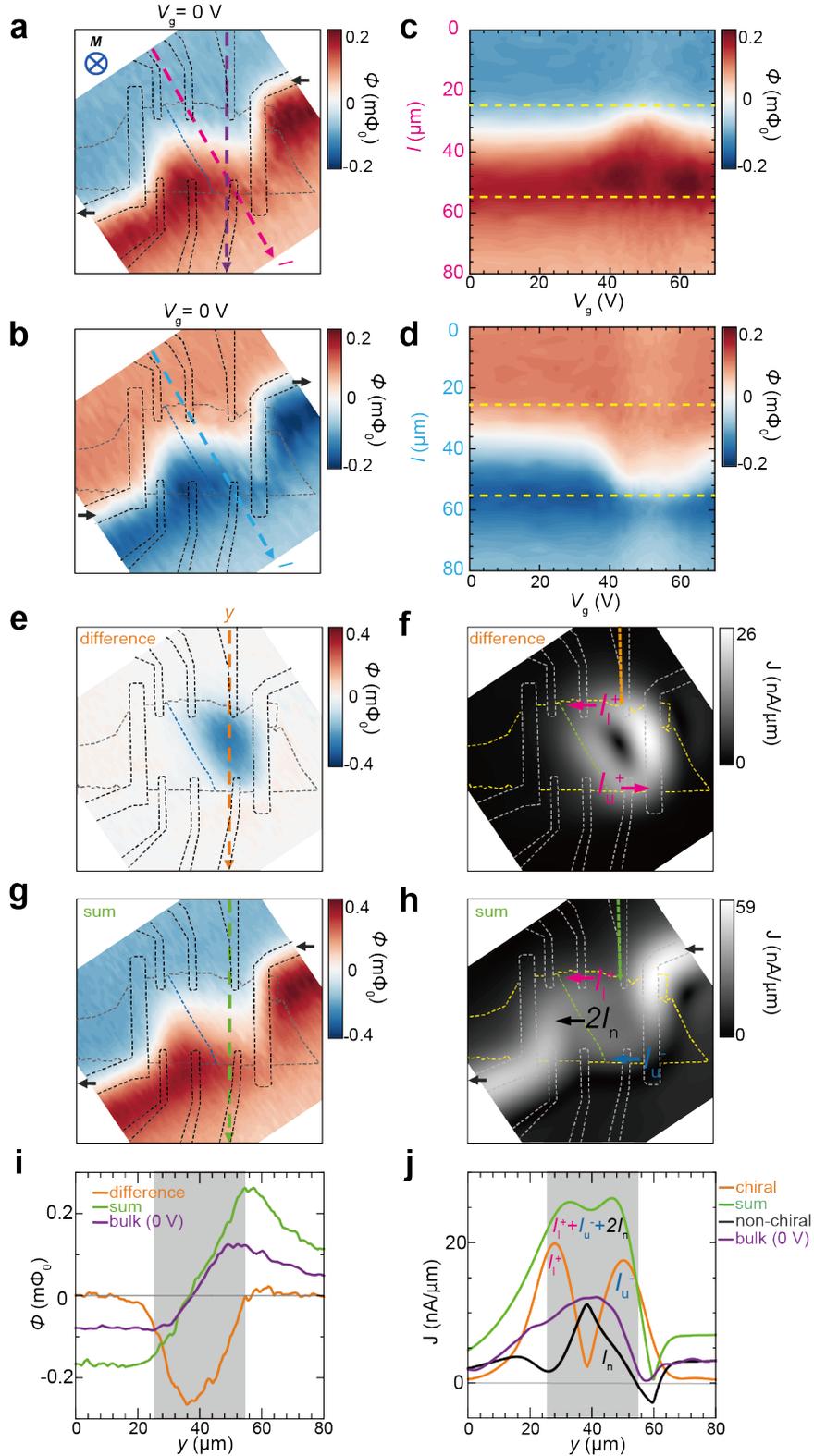

**FIG. 3 Evolution of the chiral edge current with gate tuning under opposite current bias direction. a** and **b,** Current flux images at $V_g = 0$ V with bias current on the right and left electrodes, respectively. The sample is magnetized with a -9 T field. **c** and **d,** Line scans of current flux as a function of $V_g$ along the dashed line in **a** for the two bias directions. The yellow dashed lines delineate the two edges of the device. The evolution of the current flux distribution with $V_g$

suggests that the chiral edge current appears only within the bulk magnetic exchange gap. **e,** Difference image of the two current flux images with opposite magnetization and the same bias direction (Figs. 2c and d). **f,** Current density image reconstructed from the difference current flux image in **e**. The subtraction removes the flux contribution from the non-chiral bulk current ($I_n$) while keeping one of the chiral edge currents ($I_l^+$) and reversing the $I_u^-$ to $I_u^+$ so that the difference image shows a circulation around the 7-SL clockwise. **g** and **h,** The sum current flux of Figs. 2c and d and its current density reconstruction, respectively. The sum image doubles the non-chiral contribution while adding the chiral edge currents with opposite chirality: $2I_n + I_l^+ + I_u^-$. Thus, the direction of flow in the sum image is the same for both bulk and two edge currents. **i,** Line cuts from the difference (blue) and sum (purple) current flux images across the upper and lower edges shown by the dashed lines in **e** and **g**. **j,** Line cuts from the current density $J$ images in **f** and **h** at the same position, representing the chiral (orange) and the sum (green), respectively. The non-chiral current density (black) is obtained by subtracting the chiral current density from the sum and then dividing by 2. The bulk current (purple) is obtained from the line cut of $J$ reconstructed from the current flux image at 0 V (**a**). By integrating over the width of the sample, we find that about 70% of the total current in the CNP regime is chiral edge current.

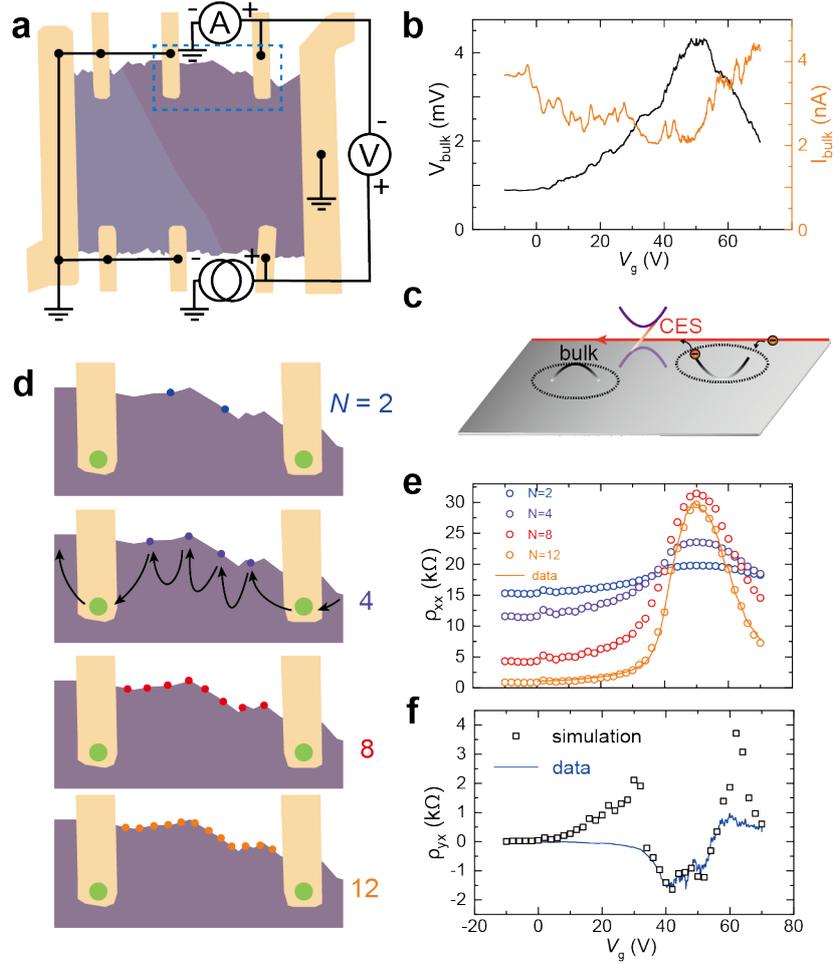

**FIG. 4 Landauer-Büttiker multi-terminal simulation of chiral edge transport with edge-bulk scattering**. **a,** Setup for measuring bulk resistance $R_{bulk}$ using electrodes on the 7-SL with all the other electrodes grounded. **b,** The voltage drop and the current of the bulk as a function of $V_g$ under the configuration shown in **a**. At the CNP, the bulk current is suppressed but $R_{bulk}$ peaks, suggesting enhanced edge-bulk scattering. **c,** Illustration of the edge-bulk scattering at the CNP. Puddles of chemical potential inhomogeneity (dashed circles) act as bulk scattering centers for chiral edge current carriers (orange balls). CES: chiral edge state. **d,** Schematics to simulate chiral edge transport between two electrodes with finite bulk-edge scattering. The bulk-edge scattering is taken into account by inserting various numbers of virtual electrodes ($N$ = 2, 4, 8 and 12). **e,** Simulated gate dependence $\rho_{xx}$ as a function of $V_g$ under different $N$. $\rho_{xx}$ increases with increasing $N$ at CNP, which suggests enhanced dissipation of chiral edge transport due to scattering with the bulk. $N$ = 12 (orange circles) overlaps with experimental $\rho_{xx}$ data at zero field (orange line). **f,** Gate dependence of simulated $\rho_{yx}$ for $N$ = 12 (black squares) and experimental $\rho_{yx}$ data at zero field (blue line), which suggests that the edge-bulk scattering suppresses the quantization of Hall resistance of QAH.